\RequirePackage{fix-cm}
\documentclass[smallextended]{svjour3}       
\smartqed  % flush right qed marks, e.g. at end of proof
\usepackage{graphicx}
\usepackage{dsfont}
\usepackage{float}
\usepackage{subfig}
\usepackage{amssymb,amsfonts}
\usepackage{amsmath}
\usepackage{color}

\journalname{Mathematics and Financial Economics}
\newcommand{\half}{\mbox{$\textstyle \frac{1}{2}$}}

\newcommand{\re}{\mbox{$\rm e$}}

\newcommand{\rd}{\mbox{$\rm d$}}
\begin{document}

\title{Coherent Chaos Interest Rate Models}


\author{Dorje~C.~Brody \and Stala~Hadjipetri}


\institute{Dorje~C.~Brody \at
              Department of Mathematics, Brunel University, Uxbridge UB8 3PH, UK \\
              \email{Dorje.Brody@brunel.ac.uk}           %  \\
           \and
           Stala~Hadjipetri \at
           Department of Mathematics, Imperial College London, London SW7 2AZ, UK \\
            \email{Stala.Hadjipetri08@imperial.ac.uk}
}

\date{\today}% Received: date / Accepted: date}

\maketitle

\begin{abstract}
The Wiener chaos approach to interest rate modelling arises from the observation that 
the pricing kernel admits a representation in terms of the conditional variance of a 
square-integrable random variable, which in turn admits a chaos expansion. When the 
expansion coefficients factorise into multiple copies of a single function, then the resulting 
interest rate model is called coherent, whereas a generic interest rate model will necessarily 
be incoherent. Coherent representations are nevertheless of fundamental importance 
because incoherent ones can always be expressed as a linear superposition of coherent 
elements. This property is exploited to derive general expressions for the pricing kernel and 
the associated bond price and short rate processes in the case of an $n$th order chaos 
model for each $n$. The pricing formulae for bond options and swaptions are obtained in 
closed forms for a number of examples. An explicit representation for the pricing kernel of 
a generic---incoherent---model is then obtained by use of the underlying coherent elements. 
Finally, finite-dimensional realisations of the coherent chaos models are investigated in detail. 
In particular, it is shown that a class of highly tractable models can be constructed having the 
characteristic feature that the discount bond price is given by a piecewise flat (simple) process. 

\end{abstract}

\section{Introduction}

For more than three decades now, interest rate modelling has been developed notably, 
generating a wide range of approaches embodying different emphases (see, e.g., James 
\& Webber 2000, Cairns 2004, Brigo \& Mercuiro 2006, Filipovic 2009, Bj\"ork 2009, Carmona 
\& Tehranchi 2010). One of the more recent approaches that has attracted some attention is 
based on the specification of the pricing kernel, from which interest rate dynamics are deduced. 
The main advantage of the pricing kernel methodology is that it allows a wide range of derivatives, 
across different asset classes, to be treated and priced in a consistent and transparent manner (cf. 
Cochrane 2005). An early example is Flesaker \& Hughston (1996, 1997, 1998), who introduced 
an approach that incorporates interest rate positivity in a canonical way. Extensions of the 
positive interest approach include Rutkowski (1997) and Jin \& Glasserman (2001). Also 
within the positive-interest context, Rogers (1997, 2004) developed a `potential approach' 
for the modelling of the pricing kernel, based on the observation that the pricing kernel 
belongs to a certain class of probabilistic potentials. 

The purpose of the present paper is to develop a new class of interest-rate models, called 
`coherent interest-rate models', within the pricing kernel formalism. Coherent interest rate 
models emerge in the context of the Wiener-chaos expansion for the pricing kernel, 
originally introduced by Hughston \& Rafailidis (2004) who made the observation that the 
class of potentials adequate for the characterisation of the pricing kernel admits a 
representation in terms of the conditional variance of a certain random variable, and 
proposed the use of the Wiener-Ito chaos expansion to model and calibrate the underlying 
random variable. The chaotic approach to interest-rate modelling was extended further in 
Brody \& Hughston (2004), where the most general forms of positive-interest arbitrage-free 
term-structure dynamics, within the context of Brownian filtration, are obtained by exploiting 
calculus on function spaces. See also Grasselli \& Hurd (2005) and Grasselli \& Tsujimoto 
(2011) for further important contributions in the `chaotoc' approach to interest rate modelling. 

In the present paper, we shall extend the chaos-based models for the pricing kernel in 
two distinct ways: (a) by working out the general representations for the chaos models at 
each chaos order; and (b) by introducing finite-dimensional realisations of chaos models based 
on function spaces. With these objectives in mind, the paper is organised as follows. In 
sections \ref{sec:2}--\ref{sec:4} we shall briefly review the background material for the 
benefit of readers less acquainted with the material, so as to make the present paper 
reasonably self-contained. Specifically, in section~\ref{sec:2} we give a brief description of 
the pricing kernel and its role in financial modelling, based on the axiomatic framework of 
Hughston \& Rafailidis (2004). Section~\ref{sec:3} summarises the argument leading to the 
conditional variance representation of Hughston \& Rafailidis (2004) (see also Bj\"ork 2007) and 
the associated chaos expansion for calibration. In section~\ref{sec:4} we explain the definition 
of the coherent chaos representation introduced in Brody \& Hughston (2004), and its role in 
interest-rate modelling. 

In section~\ref{sec:5} we introduce the notion of an $n^{\rm th}$-order coherent chaos model, 
and derive the general representation for the pricing kernel, the short rate, the bond price, 
and the risk premium in this model. Explicit examples of derivative pricing formulae are then 
obtained in section~\ref{sec:6} for coherent interest rate models. Specifically, bond option and 
swaption prices are obtained in closed form. Coherent chaos models are important because 
they form the building block for general interest rate models. Specifically, a generic interest 
rate model can be expressed in the form of a linear superposition of the underlying coherent 
components (in the sense explained in section~\ref{sec:4}), and thus are incoherent. By exploiting 
the linearity structure we are able to derive a generic expression in section~\ref{sec:7} for the 
pricing kernel for $n^{\rm th}$-order incoherent chaos models. 

In section~\ref{sec:8} we shift the gear slightly by investigating finite-dimensional realisations 
of the coherent chaos models. We note that a chaos coefficient is given by an element of an 
infinite-dimensional Hilbert space of square-integrable functions. When this coefficient is replaced 
by the square-root of a \textit{finite} number of Dirac delta functions, then the resulting system 
can in effect be treated in a finite-dimensional Hilbert space. The corresponding interest rate models 
are `simple' in the sense that the pricing kernel and the bond price processes are given by 
piecewise step functions, where step sizes are sampled from nonlinear functions of independent 
Gaussian random variables. The idea of the finite-dimensional realisation example is that it can be 
used to fit a finite-number of initial bond prices, without relying on interpolation methods. Although 
the resulting models are rather elementary, we found them to be nevertheless of some interests. 
We conclude in section~\ref{sec:9} with a brief discussion and further comments.

\section{The pricing kernel approach to interest rate modelling}
\label{sec:2} 

The pricing kernel approach to valuation and risk management provides perhaps the most 
direct route towards deriving various familiar results in financial modelling. It also provides an 
insight into understanding the relation between risk, risk aversion, and return arising from 
risky investments when asset prices can jump---an idea that is perhaps difficult to grasp from other 
approaches available (Brody \textit{et al}. 2012). For these reasons here we shall adapt the pricing 
kernel approach to interest rate modelling. We shall begin in this section with a brief review of the 
idea of a pricing kernel. In particular, we find it convenient to follow the axiomatic approach introduced 
in Hughston \& Rafailidis (2004). 

The axioms listed below are neither minimum nor unique (see also Brody \& Hughston 2004, 
Rogers 2004, Hughston \& Mina 2012), however, we find that the approach taken in Hughston 
\& Rafailidis (2004) leads to the desired representation of the pricing kernel needed for our 
purpose in the most expedient manner, and thus we shall take this as our starting point. 

We proceed as follows. We model the economy with a fixed probability space $(\Omega, 
\mathcal{F},{\mathbb{P}})$, where the probability measure $\mathbb{P}$ is the physical 
probability measure. We equip this space with the standard augmented filtration 
$\{\mathcal{F}_t\}_{0\leq t<\infty}$ generated by a system of one or more independent Wiener 
processes $\{W_t^\alpha\}_{0\leq t<\infty}$, $\alpha= 1,\ldots, k$. We also assume that asset 
prices are continuous semimartingales on $(\Omega, \mathcal{F},\mathbb{P})$, which will enable 
us to utilise various standard results of stochastic calculus. With this setup in mind, Hughston \& 
Rafailidis (2004) characterises the absence of arbitrage in the economy by assuming 
the existence of a pricing kernel, a strictly positive Ito process $\{\pi_t\}_{t \geq 0}$ on 
$(\Omega, \mathcal{F},\mathbb{P})$, such that the following set of axioms hold:
\begin{itemize} 
\item[(a)] There exists a strictly increasing absolutely continuous asset with price process 
$\{B_t\}_{t \geq 0}$ that represents the money market account.
\item[(b)] Given any asset with price process $\{S_t\}_{t \geq 0}$ and with 
$\mathcal{F}_t$-adapted dividend rate process $\{D_t\}_{t \geq 0}$, the process $\{M_t\}$ 
defined by
\begin{eqnarray}
M_t = \pi_t S_t + \int_0^t \pi_s D_s \rd s
\end{eqnarray}
is a ${\mathbb P}$-martingale.
\item[(c)] There exists an asset---a floating rate note---that offers a continuous dividend rate such 
that its value remains constant over time. 
\item[(d)] There exists a system of discount bonds $\{P_{tT}\}_{0 \leq t \leq T\leq\infty}$ with the 
transversal property that 
\begin{eqnarray}\label{lim P}
\lim_{T \rightarrow \infty} P_{tT} = 0, \quad \mathbb{P}-\textrm{a.s.}
\end{eqnarray}
\end{itemize} 

From axiom (a), we deduce the existence of an  $\mathcal{F}_t$-adapted short rate process 
$\{r_t\}_{t\geq 0}$ such that $r_t>0$ for all $t \geq 0$ and that 
\begin{eqnarray} \label{B sde}
\rd B_t = r_t B_t \rd t . 
\end{eqnarray}
Since the money market account is an asset that pays no dividend, it follows from axiom (b) that 
the process $\{\rho_t\}_{t\geq 0}$ defined by 
\begin{eqnarray}\label{lambda}
\rho_t= \pi_t B_t
\end{eqnarray} 
is a martingale. Note that at most one process $B_t$ satisfying (a) and (b) can exist. Due to the 
fact that $\pi_t>0$ and $B_t >0$, $\rho_t$ is also strictly positive for all $t \geq 0$. Therefore, 
$\{\rho_t\}$ is a positive martingale and we can write
\begin{eqnarray}\label{Lambda sde}
\rd \rho_t = - \rho_t \lambda_t \rd W_t 
\end{eqnarray}
for some $\mathcal{F}_t$-adapted vector-valued process $\{\lambda_t\}_{t\geq 0}$. Here and 
in what follows, for simplicity of notation we shall write $\lambda_t \rd W_t$ to denote the vector 
inner product $\sum_{\alpha = 1}^k \lambda_t^\alpha \rd W_t^\alpha$. As a consequence of 
(\ref{B sde}), (\ref{lambda}), and (\ref{Lambda sde}), the dynamical equation satisfied by the 
pricing kernel takes the form 
\begin{eqnarray} \label{pi sde}
\rd \pi_t = -r_t \pi_t \rd t - \lambda_t \pi_t \rd W_t,
\end{eqnarray}
or equivalently we can write 
\begin{eqnarray} \label{pi}
\pi_t = \exp\left(-\int_0^t r_s {\rm d}s - \int_0^t \lambda_s {\rm d} W_s-\half \int_0^t 
\lambda_s^2 {\rm d}s\right) .
\end{eqnarray}
Since the pricing kernel is a product of a martingale and a strictly decreasing process, it follows 
that $\{\pi_t\}$ is a supermartingale satisfying ${\mathbb E}_t[\pi_T]\leq\pi_t$. 

Consider now a system of default-free discount (zero coupon) bonds. We assume, in accordance 
with axiom (d), that the economy supports such a system of discount bonds over all time horizons. 
We write $P_{tT}$ for the value at time $t$ of a default-free $T$-maturity discount bond that pays 
one unit of currency at maturity $T$. Since a discount bond pays no dividend, we deduce from 
axiom (b) that for each maturity $T$ the process $\{\pi_t P_{tT}\}_{0 \leq t \leq T}$ is a martingale, which 
we shall denote by $\{M_{tT}\}_{0 \leq t \leq T}$. It then follows from equation (\ref{lambda}) that  
\begin{eqnarray} \label{p=L over B}
P_{tT} = \frac{M_{tT} B_t}{\rho_t} . 
\end{eqnarray}
Since $\{M_{tT}\}$ is a parametric family of martingales, there exists a vector-valued family of 
processes $\{\sigma_{tT}\}_{0 \leq t \leq T}$ such that we can write 
\begin{eqnarray} \label{L sde}
\rd M_{tT} = M_{tT} (\sigma_{tT} - \lambda_t) \rd W_t.
\end{eqnarray}
Using Ito's product and quotient rules, it follows from equations (\ref{B sde}), (\ref{Lambda sde}), 
(\ref{p=L over B}) and (\ref{L sde}) that the dynamical equation satisfied by the discount-bond system 
is given by
\begin{eqnarray} \label{P sde}
\rd P_{tT} = P_{tT} (r_t + \lambda_t \sigma_{tT})\rd t + P_{tT} \sigma_{tT} \rd W_t,
\end{eqnarray}
which can alternatively be written in the form 
\begin{eqnarray}
P_{tT} = P_{0T} B_t \exp \left( \int _0^t \sigma_{sT} \left( \rd W_s
+ \lambda_s \, \rd s \right) - \half \int _0^t \sigma_{sT}^2 \,
\rd s \right) . \label{4-2.4}
\end{eqnarray}
We therefore recognise the process $\{\sigma_{tT}\}_{0 \leq t \leq T}$ for a fixed $T$ as the 
$T$-maturity discount bond volatility, while $\{\lambda_t\}_{t\geq0}$ measures the excess rate of 
return above the short rate arising from risky investments per volatility.

As remarked above, the pricing kernel approach also provides an effective way for the 
valuation of a contingent claim. Thus, for example, if $H_T$ is the payout of a derivative 
at time $T$, then on account of axiom (b) we find that the value of the derivative at time 
$t$ is given by 
\begin{eqnarray}
H_t = \frac{{\mathbb E}_t[\pi_T H_T]}{\pi_t} . 
\label{eq:12} 
\end{eqnarray}
In particular, for a unit cash flow $H_T=1$ we obtain the bond pricing formula 
\begin{eqnarray}
P_{tT} = \frac{{\mathbb E}_t[\pi_T]}{\pi_t} . 
\label{bond}
\end{eqnarray}
If we now take the limit $T\to\infty$ in (\ref{bond}), and make use of (\ref{lim P}), we find that 
\begin{eqnarray} 
\lim_{T\rightarrow \infty} {\mathbb E}_t[\pi_T]= 0, \quad \mathbb{P} \; \mbox{- a.s.}
\end{eqnarray}
This shows that the pricing kernel is a potential, i.e. a right-continuous positive supermartingale 
whose expectation value vanishes asymptotically as $T \rightarrow \infty$. More specifically, 
$\{\pi_t\}$ is a type-D potential (in the language of Meyer 1966). This observation led Rogers 
(1997) to introduce the so-called potential approach to interest rate modelling (see also Rogers 
2004, Bj\"ork 2007). 

The specification of $\{\pi_t\}$ therefore leads on the one hand to the bond price dynamics as 
well as the associated short rate process, while on the other hand to the pricing of general 
contingent claims. It is for these reasons that we prefer to model $\{\pi_t\}$ directly. One such 
approach, as indicated above, is to consider certain types of potentials for the basis of interest 
rate modelling; another approach, which we shall consider here, is to employ the Wiener chaos 
expansion technique to build interest rate models.

\section{Conditional variance and Wiener chaos expansion}
\label{sec:3}

To proceed we shall follow the observation made in Hughston \& Rafailidis (2004) that the pricing 
kernel can be expressed in the form of a conditional variance of an ${\mathcal F}_\infty$-measurable 
square-integrable random variable. The idea is as follows. We write (\ref{pi sde}) in the integral form 
\begin{eqnarray}
\pi_T - \pi_t = - \int_t^T r_s \pi_s \rd s - \int_t^T \lambda_s \pi_s \rd W_s , 
\label{eq:15}
\end{eqnarray} 
and take the ${\mathcal F}_t$-conditional expectation on each side of (\ref{eq:15}) to obtain 
\begin{eqnarray}
\pi_t = {\mathbb E}_t[\pi_T] + {\mathbb E}_t\left[ \int_t^T r_s \pi_s \rd s \right] . 
\label{eq:16}
\end{eqnarray} 
Now taking the limit $T\to\infty$ we deduce the following implicit relation for the pricing kernel: 
\begin{eqnarray} \label{pi exp int}
\pi_t = \mathbb{E}_t\left[ \int_t^\infty r_s \pi_s \rd s\right].
\end{eqnarray}
Note that since $\{\pi_t\}$ and $\{r_t\}$ are both $\mathcal{F}_t$-adapted, the vector valued process 
$\{\eta_t\}_{t \geq 0}$ defined by the relation 
\begin{eqnarray}
\eta_t^2 = \sum_{\alpha=1}^k \eta_t^\alpha \eta_t^\alpha = r_t \pi_t
\end{eqnarray}
is also $\mathcal{F}_t$-adapted. Evidently, the vector $\eta_t$ for each $t\geq0$ is unique only up 
to SO($k$)-rotational degrees of freedom. For any given representative element $\{\eta_t\}$ of this 
equivalence class we define an ${\mathcal F}_t$-martingale $\{X_t\}$ according to the following 
prescription: 
\begin{eqnarray} 
X_t = \int_0^t \eta_s \rd W_s . 
\end{eqnarray}
Then we have 
\begin{eqnarray} 
X_\infty - X_t = \int_t^\infty \eta_s \rd W_s , 
\end{eqnarray}
from which, on account of the conditional Wiener-Ito isometry we deduce that 
\begin{eqnarray} 
\pi_t = \mathbb{E}_t\left[ \left( X_\infty - \mathbb{E}_t[X_\infty]\right)^2\right].
\label{21}
\end{eqnarray}
We shall refer to this identity as the conditional-variance representation for the pricing kernel. 

The foregoing analysis shows that to model the pricing kernel, it suffices to model the random 
variable $X_\infty$ that is an element of the Hilbert space $\mathcal{L}^2(\Omega)$ of square 
integrable random variables---more precisely, an element of the subspace of $\mathcal{L}^2(\Omega)$ 
consisting of $\mathcal{F}_\infty$-measurable random variables. The proposal of Hughston \& 
Rafailidis (2004) is to employ the Wiener chaos expansion to `parameterise' the random variable 
$X_\infty$, work out the pricing kernel $\{\pi_t\}$, and use the result to obtain pricing formulae for 
various derivatives; which in turn allows one to calibrate the functional parameters in the chaos 
expansion. Specifically, the random variable $X_\infty$ admits a unique expansion of the form 
\begin{eqnarray}
X_\infty = \phi + \int_0^\infty \phi(s_1) \rd W_{s_1} + \int_0^\infty \left(
\int_0^{s_1} \phi(s_1, s_2) \rd W_{s_2} \right) \rd W_{s_1} +
\cdots, \label{eq:22}
\end{eqnarray}
where $\phi={\mathbb E}[X_\infty]$, $\phi_s=\phi(s)\in{\mathcal L}^2({\mathds R}_+)$ is a 
square-integrable function of one variable, $\phi_{ss'}=\phi(s,s')\in{\mathcal L}^2({\mathds R}_+) 
\otimes{\mathcal L}^2({\mathds R}_+)$ is a (symmetric) square-integrable function of two variables, 
and so on. 

We remark incidentally that the idea of a chaos expansion was introduced by Wiener in his seminal 
paper titled ``Homogeneous chaos'' (Wiener 1938); whereas the representation (\ref{eq:22}) in terms 
of the stochastic integral is due to Ito (1951). We refer to Ikeda \& Watanabe (1989), Nualart (1995), 
Janson (1997), Malliavin (1997), {\O}ksendal (1997), and Di~Nunno \textit{et al}. (2009) for further 
discussion of the chaos expansion technique and its role in stochastic analysis. 

For interest-rate modelling, therefore, we see that by substituting (\ref{eq:22}) in (\ref{21}), the pricing 
kernel is parameterised by a vector ${\boldsymbol\Phi}$ of a set of deterministic quantities: 
\begin{eqnarray}
{\boldsymbol\Phi} = \big( \phi, \ \phi(s), \ \phi(s,s_1), \ \phi(s,s_1,s_2),\ \cdots \big) ,
\end{eqnarray}
where ${\boldsymbol\Phi}$ is itself an element of a Fock space ${\mathfrak F}$ of the direct sum of 
the Hilbert spaces of square-integrable functions. The elements of ${\boldsymbol\Phi}$ are called 
the {\it Wiener chaos coefficients}, or {\it Wiener-Ito chaos coefficients}, and these coefficients fully 
characterise the information in $X_\infty$. Since $X_\infty$ determines the pricing kernel $\{\pi_t\}$, 
which in turn determines the bond price process $\{P_{tT}\}$, each interest rate model can be viewed 
as depending on the specification of its Wiener-Ito chaos coefficients.

\section{Coherent chaos expansion} 
\label{sec:4} 

There is a special class of vectors in the Fock space ${\mathfrak F}$ called `coherent vectors' that 
admit a number of desirable characteristics. Let us consider the tensor product ${\mathcal H}^{(n)} = 
{\mathcal L}^2({\mathds R}_+)\otimes{\mathcal L}^2({\mathds R}_+)\otimes\cdots\otimes
{\mathcal L}^2({\mathds R}_+)$ of $n$ copies of Hilbert spaces. A generic element of 
${\mathcal H}^{(n)}$ may be written in the form $\phi(s_1,s_2,\cdots,s_n)$; whereas a coherent vector 
of ${\mathfrak F}$ is generated by a map from an element of ${\mathcal H}^{(1)}$ to an element of 
${\mathcal H}^{(n)}$. Specifically, given $\phi(s)\in{\mathcal H}^{(1)}={\mathcal L}^2({\mathds R}_+)$, 
we consider an element of ${\mathcal H}^{(n)}$ of the degenerate form 
\begin{eqnarray}
\phi(s_1,s_2,\cdots,s_n) = \phi(s_1)\phi(s_2) \ldots \phi(s_n) . 
\end{eqnarray}
The importance of coherent vectors is that the totality of such vectors in ${\mathcal H}^{(n)}$ 
constitutes a resolution of the identity, i.e. these vectors are in general not orthogonal but nevertheless 
are complete. Therefore, an arbitrary element of ${\mathcal H}^{(n)}$ can be expressed in the form of 
a linear combination of (possibly uncountably many) coherent vectors. 

More generally, given an element $\phi(s)\in{\mathcal H}^{(1)}$ we can generate a coherent Fock 
vector of the form 
\begin{eqnarray}
{\boldsymbol\Phi}_\phi = \big( 1 , \phi(s), \phi(s_1)\phi(s_2), \cdots, \phi(s_1)\phi(s_2) \ldots \phi(s_n) \big) . 
\end{eqnarray} 
Then an arbitrary element of ${\mathfrak F}$ can likewise be expressed as a linear combination of 
coherent vectors. The significance of the completeness of coherent vectors for the chaos expansion, 
noted in Brody \& Hughston (2004), is as follows. We observe first that if ${\boldsymbol\Phi}_\phi$ 
is coherent, then the associated random variable $X_\infty^\phi$ arising from the chaos expansion 
(\ref{eq:22}) takes the form 
\begin{eqnarray} \label{eq:26}
X_\infty^\phi = \sum_{n=0}^\infty \left(\int\limits_0^\infty \int\limits_0^{s_1} \cdots\!\! \int\limits_0^{s_{n-1}} 
\phi(s_1)\phi(s_2) \cdots \phi(s_n) \rd W_{s_n}\cdots \rd W_{s_2} \rd W_{s_1}\right),
\end{eqnarray}
where the $n=0$ term is assumed to be unity. We now make use of the following identity due to 
Ito (1951): 
\begin{eqnarray}
\int\limits_0^T  \int\limits_0^{s_1} \cdots\!\! \int\limits_0^{s_{n-1}}\! \phi(s_1) \phi(s_2) \cdots 
\phi(s_{n}) \rd W_{s_{n}}\!\!\cdots \rd W_{s_1} = \frac{Q_T^{n/2}}{n!} H_n\left( \frac{R_T}
{Q_T^{1/2}}\right),
\label{eq:27}
\end{eqnarray}
where
\begin{eqnarray}
R_t = \int_0^t \phi(u) \rd W_u \quad {\rm and} \quad Q_t = \int_0^t \phi^2(u) \rd u, 
\label{eq:28}
\end{eqnarray}
and where 
\begin{eqnarray}
H_n(x)= n! \sum_{k=0}^{\left\lfloor n/2\right\rfloor} \frac{(-1)^k \; x^{n-2k}}{k!\; (n-2k)! \;2^k} 
\label{eq:29}
\end{eqnarray}
denotes the Hermite polynomials, satisfying the generating function relation 
\begin{eqnarray}
\exp\left( tx - \half t^2\right) = \sum_{n=0}^\infty \frac{t^n}{n!} H_n(x). 
\label{eq:30} 
\end{eqnarray}
The role of Hermite polynomials in relation to Gaussian random variables is well known (see, 
e.g., Schoutens 2000). Here we merely remark the property that if $Y$ and $Z$ are standard 
normal random variables, then 
\begin{eqnarray} \label{orth}
\mathbb{E}[H_n(Y)H_m(Z)] = \left(\mathbb{E}[YZ]\right)^n \; \delta_{mn},  
\label{eq:31}
\end{eqnarray}
which follows from the fact that Hermite polynomials are orthogonal with respect to the standard 
normal density function. 

By comparing (\ref{eq:27}) and (\ref{eq:30}) we thus deduce that 
\begin{eqnarray}
X_\infty^\phi = \exp\left( \int_0^\infty \phi(s)\rd W_{s} - \half \int_0^\infty \phi^2(s) \rd s \right) . 
 \label{eq:32}
\end{eqnarray}
On account of the completeness of the coherent vectors and linearity, therefore, it follows that an 
arbitrary ${\mathcal F}_\infty$-measurable square-integrable random variable $X_\infty$ admits a 
simple representation 
\begin{eqnarray}
X_\infty = \sum_j c_j \exp\left( \int_0^\infty \phi_j(s)\rd W_{s} - \half \int_0^\infty \phi_j^2(s) \rd s \right) , 
\label{eq:33}
\end{eqnarray}
where $\{c_j\}$ are constants satisfying $\sum_jc_j^2<\infty$, where $\phi_j(s)\in{\mathcal L}^2({\mathds R}_+)$ 
for each $j$ is a deterministic square-integrable function, and where the summation in (\ref{eq:33}) 
is formal and may be replaced by an appropriate integration in the uncountable case. It should be 
evident that an analogous result holds more generally for an arbitrary ${\mathcal F}_t$-measurable 
square-integrable random variable---that is, any such random variable can be expressed in terms 
of a linear combination of log-normally distributed random variables. Putting the matter differently, 
log-normal random variables are dense in 
the space of square-integrable random variables. This remarkable fact was applied in Brody \& Hughston (2004) to identify the general 
expressions for the pricing kernel and other quantities (such as the bond price, risk premium, and the 
various interest rates) in the Brownian-based setting, owing to the fact that the conditional variance of 
$X_\infty$ in (\ref{eq:33}) is easily calculated in closed form.

\section{Coherent chaos interest-rate models} 
\label{sec:5} 

The point of departure in the present investigation is to examine more closely the $n^{\rm th}$-order 
chaos models for each $n$. That is to say, we are concerned with interest-rate models that arise from 
the chaos expansion of the form 
\begin{eqnarray}
X_\infty^{(n)} = \int_0^\infty \int_0^{s_1} \cdots \int_0^{s_{n-1}} \phi(s_1,s_2,\ldots,s_n) \, 
\rd W_{s_n}\cdots \rd W_{s_2} \rd W_{s_1} 
 \label{eq:34}
\end{eqnarray}
for each $n$, where $\phi(s_1,s_2,\ldots,s_n)\in{\mathcal H}^{(n)}$. As indicated above, any such 
function $\phi(s_1,s_2,\ldots,s_n)$ can be expressed as a linear combination of (possibly uncountable) 
coherent functions of the form $\phi(s_1)\phi(s_2)\cdots\phi(s_n)$. Our strategy therefore is to first 
work out the interest rate model arising from such a coherent element of ${\mathcal H}^{(n)}$, which we shall refer to as the $n^{\rm th}$-order coherent 
chaos model, and then consider their linear combinations for more general 
$n^{\rm th}$-order chaos models for the pricing kernel. 

To proceed, we recall that from (\ref{eq:27}) it follows that for a coherent element we have 
\begin{eqnarray}
X_\infty^{(n)} = \frac{Q_\infty^{n/2}}{n!} H_n\left( \frac{R_\infty}{Q_\infty^{1/2}}\right) 
\quad {\rm and} \quad 
X_t^{(n)} = \frac{Q_t^{n/2}}{n!} H_n\left( \frac{R_t}{Q_t^{1/2}}\right) , 
\label{eq:35}
\end{eqnarray}
where we have written $X_t^{(n)}={\mathbb E}_t[X_\infty^{(n)}]$. Note that the 
argument $R_t/Q_t^{1/2}$ appearing here in the Hermite polynomial is, for 
each $t$, a standard normally distributed random variable; hence on account 
of (\ref{orth}) we find that the martingales $\{X_t^{(n)}\}_{0 \leq t \leq \infty}$ 
are mutually orthogonal. This property will be exploited in a calculation below 
when we analyse the more general `incoherent' models. 

Let us now turn to the determination of the pricing kernel: 
\begin{eqnarray} 
\pi_t^{(n)} = \mathbb{E}_t \left[ \left(X_{\infty}^{(n)}\right)^2 \right] - \left( X_t^{(n)}\right)^2 .
\label{eq:36}
\end{eqnarray}
To analyse these expressions, we make use of the following product identity 
for a pair of Hermite polynomials of different order 
(see, e.g., Janson 1997, p.28):
\begin{eqnarray}
H_n(x)H_m(x) = \sum_{k=0}^{m \wedge n} \left(\begin{array}{c}
m\\k \end{array} \right) \left(\begin{array}{c}
n\\k \end{array} \right) k! \; H_{m+n-2k} (x) , 
\end{eqnarray}
where $m \wedge n = \min(m,n)$. Setting $m=n$ we obtain
\begin{eqnarray} 
H_n^2 (x) = \sum_{k=0}^n \left(\begin{array}{c} n\\k \end{array} \right) \left(\begin{array}{c} n\\k
\end{array} \right) k! \; H_{2(n-k)} (x).
\label{eq:38} 
\end{eqnarray}
Since, from (\ref{eq:29}), we have 
\begin{eqnarray} 
X_{t}^{(n)}= \sum_{k=0}^{\left\lfloor n/2\right\rfloor} \frac{(-1)^k \; 
R_t^{n-2k}\; Q_t^k}{k!\; (n-2k)! \;2^k}, 
\label{eq:39}
\end{eqnarray}
it follows that by squaring $X_t^{(n)}$ of (\ref{eq:35}) and making use of (\ref{eq:38}), we 
deduce that 
\begin{eqnarray} 
\left( X_t^{(n)}\right)^2 = \sum_{k=0}^n  \frac{Q_t^k \; [2(n-k)]!\;X_{t}^{(2n-2k)}}{k! \; [(n-k)!]^2}.
\label{eq:40}
\end{eqnarray}
Taking the limit $t \to \infty$ in (\ref{eq:40}) and substituting the result in 
(\ref{eq:36}) we obtain 
\begin{eqnarray}
\pi_t^{(n)} = \sum_{k=0}^n  \frac{[2(n-k)]! \; (1- Q_t^k) \; X_{t}^{(2n-2k)} }{k!\; [(n-k)!]^2}, 
\label{eq:41}
\end{eqnarray}
where we have unit-normalised the function $\phi(s)$ so that $Q_\infty=1$. This is the desired 
expression for the pricing kernel associated with $n^{\rm th}$-order coherent chaos models. We 
observe from (\ref{eq:35}) that $\{\pi_t^{(n)} \}$ in this case is given by a polynomial of order $2n-2$ 
of the Gaussian process $\{R_t\}$. 

Having obtained the expression for the pricing kernel we are in the position to determine the 
representations for the various quantities of interest. 
To this end, let us first derive expressions for the short rate and risk premium processes in this 
model. Specifically, on account of (\ref{pi sde}) we observe that these processes are given 
by the drift and the volatility of the pricing kernel. A short calculation then shows that 
\begin{eqnarray} 
r_{t}^{(n)} = \frac{\phi^2(t)}{\pi_t^{(n)}} \; \sum_{k=0}^{n} 
\frac{[2(n-k)]! \; Q^{k-1}_t \; X_t^{(2n-2k)}}{(k-1)! \; [(n-k)!]^2}
\label{eq:42}
\end{eqnarray}
and that 
\begin{eqnarray}
\lambda_{t}^{(n)} = -\frac{\phi(t)}{\pi_t^{(n)}} \sum_{k=0}^n 
\frac{(1-Q^{k}_t)\; [2(n-k)]!}{k! \; [(n-k)!]^2}X_t^{(2n-2k-1)}, 
\label{eq:43}
\end{eqnarray}
where we have made use of the relation 
\begin{eqnarray}
\rd X_{t}^{(2n-2k)} = X_{t}^{(2n-2k-1)} \phi(t) \rd W_t 
\label{eq:44}
\end{eqnarray}
in obtaining (\ref{eq:43}). 

As for the discount bonds, it follows from (\ref{bond}) that the bond price 
process takes the form of a 
ratio of polynomials of the Gaussian process $\{R_t\}$:
\begin{eqnarray}
P_{tT}^{(n)} = \frac{\sum\limits_{k=0}^n \frac{[2(n-k)]! \; (1 - Q_T^k) \; X_{t}^{(2n-2k)}}
{k! \; [(n-k)!]^2}} 
{\sum\limits_{k=0}^n \frac{[2(n-k)]! \; (1 - Q_t^k)\;X_{t}^{(2n-2k)}}{k!\; [(n-k)!]^2}},
\end{eqnarray}
with the initial term structure 
\begin{eqnarray} \label{P0t}
P_{0T}^{(n)}= 1-Q_T^n.
\label{initial bond}
\end{eqnarray}
This shows, in particular, that in the case of a single-factor setup, coherent 
chaos models are fully characterised by the initial term structure. Such a strong 
constraint, of course, is expected in the case of a single factor model for which 
the only model `parameter' is a deterministic scalar function $\phi(t)$, and it is 
indeed natural that this functional degrees of freedom should be fixed 
unambiguously by the 
initial yield curve. On the other hand, the general cases are obtained 
merely by taking linear superpositions of the coherent chaos models in an 
appropriate manner. Before we examine these general cases, let us first 
work out a number of derivative-pricing formulae in the case of 
single-factor coherent chaos models.

\section{Derivative pricing formulae in coherent chaos models} 
\label{sec:6}

We recall expression (\ref{eq:12}) for the pricing of a generic derivative. 
The purpose of this section is to derive bond option and swaption pricing 
formulae in a number of example models. Specifically, we shall examine the 
pricing of bond options and swaptions for the second and third order coherent 
chaos models. 

\subsection{Bond option pricing for second-order coherent chaos models} 

In the case of a second-order coherent chaos model, it follows from (\ref{eq:41}) that the pricing 
kernel is given by the following quadratic function of a single Gaussian state variable $R_t$: 
\begin{eqnarray}
\pi_{t}^{(2)}=(1-Q_t)(R_t^2-Q_t)+\half(1-Q_t^2). 
\label{eq:47}
\end{eqnarray}
The bond price process can then be written as the following fraction:
\begin{eqnarray}
P_{tT}^{(2)}=\frac{(1-Q_T)(R_t^2 - Q_t)+\frac{1}{2}(1-Q_T^2)}
{(1-Q_t)(R_t^2-Q_t)+\frac{1}{2}(1-Q_t^2)}.
\label{eq:48}
\end{eqnarray}

We shall make use of (\ref{eq:47}) and (\ref{eq:48}) to find the pricing formula for a European-style 
call option on the discount bond. In particular, we let $t$ be the option maturity on a bond that matures 
at $T\geq t$, and $K$ be the option strike. Then bearing in mind that $\pi_0^{(2)}=\frac{1}{2}$ under the 
convention $Q_\infty=1$ that we have chosen here, the initial price $C_0$ of the option is given by 
\begin{eqnarray} 
C_0^{(2)}(t,T,K) 
&=& 2\,\mathbb{E}\left[\pi_t^{(2)} \left(P_{tT}^{(2)}-K\right)^+\right] \nonumber \\ &=& 
2 \, \mathbb{E}\left[\Big(R_t^2 \left[(1- Q_T)-K(1-Q_t)\right]- Q_t(1-Q_T)  \right. \nonumber \\ && 
\left. \left. \qquad + \half (1-Q_T^2) -\half K \left(1- Q_t\right)^2\right)^+\right]\nonumber \\ &=& 
2 \, \mathbb{E}\left[\left(A Z_t^2 + B\right)^+\right], 
\label{eq:49}
\end{eqnarray}
where $Z_t = R_t/\sqrt{Q_t}$ is a standard normal random variable, and where 
\begin{eqnarray}
A = Q_t \left[(1- Q_T)-K(1-Q_t)\right], 
\end{eqnarray} 
and 
\begin{eqnarray} 
B = \half \left[(1- Q_T^2) - K (1-Q_t^2)\right]
- Q_t \left[(1-Q_T)-K (1-Q_t)\right].
\end{eqnarray}
The problem of pricing the call option in this model thus reduces to finding the roots of a quadratic 
equation $Az^2+B=0$. Letting $z_1,z_2$ denote the roots: 
\begin{eqnarray}
z_{1}= -\sqrt{-\frac{B}{A}} \quad {\rm and} \quad z_{2}= +\sqrt{-\frac{B}{A}},
\end{eqnarray}
we must consider different scenarios depending on the signatures of the coefficients 
$A$ and $B$. We shall proceed to examine this case by case. 

\begin{itemize} 
\item[(i)] If $A=0$ then $B = \frac{1}{2} (1-Q_T)(Q_T - Q_t)>0$ since $Q_T >Q_t$, so the option is always in the 
money and we have 
\begin{eqnarray}
C_0^{(2)} = P_{0T}^{(2)}-K P_{0t}^{(2)} = (1- Q_T)(Q_T-Q_t), 
\end{eqnarray}
which is just $2B$, as is evident from (\ref{eq:49}). 
\item[(ii)] If $A>0$ then $B > 0$. This follows from the fact that $A>0$ implies $K< (1-Q_T)/(1-Q_t)$. Since $B= \frac{1}{2}(1-Q_T^2) - Q_t (1-Q_T) - \frac{1}{2}K(1-Q_t)^2$, the bound on $K$ implies $B>\frac{1}{2} (1+Q_T)(Q_T - Q_t)$ but since $Q_T>Q_t$, we have $B>0$. In this case, $Az^2+B$ is 
always positive, so again the call option will always expire in the money. The price of such an option 
is thus 
\begin{eqnarray} 
C_0^{(2)} &=& \sqrt{\frac{2}{\pi}} \int_{-\infty}^{\infty} (Az^2+B) \; {\rm e}^{-\frac{1}{2}z^2} \rd z
\nonumber \\ &=& P_{0T}^{(2)}-KP_{0t}^{(2)} \nonumber \\ &=& (1- Q_T^2)-K(1-Q_t^2), 
\label{C_0i} 
\end{eqnarray} 
which is just $2(A+B)$. 
\item[(iii)] If $A<0$ and $B \leq 0$, then the payoff is never positive, resulting in a worthless option 
whose value is zero. 
\item[(iv)] The only nontrivial case is when we have that $A<0$ and $B> 0$. Then the quadratic 
polynomial is positive over the interval $(z_1,z_2)$, and we have 
\begin{eqnarray}
C_0^{(2)} &=& \sqrt{\frac{2}{\pi}} \int_{z_{1}}^{z_{2}}  \left(A z^2  + B\right) \; {\re}^{-\frac{1}{2}z^2}  \rd z 
\nonumber \\ &=&  (P_{0T}^{(2)}-K P_{0t}^{(2)})\left[1- 2 N(z_{1})\right]+ 4 A z_1 \rho(z_1)Q_t, 
\end{eqnarray}
where $N(z)$ denotes the standard cumulative normal distribution function, and $\rho(z)$ the associated 
density function. A short calculation also shows that the option delta that gives the position on the 
underlying bond required to hedge the option, in this case, is given by 
\begin{eqnarray}
\Delta = 1- 2 N(z_1) + \frac{1}{2 A z_1}\, \rho(z_1) (P_{0T}- K P_{0t})(4A z_1 - 1) . 
\end{eqnarray}
\end{itemize} 

\begin{figure}[t!]
\centering
\subfloat{\includegraphics[trim=1cm 0cm 0cm 0cm, scale=0.31]{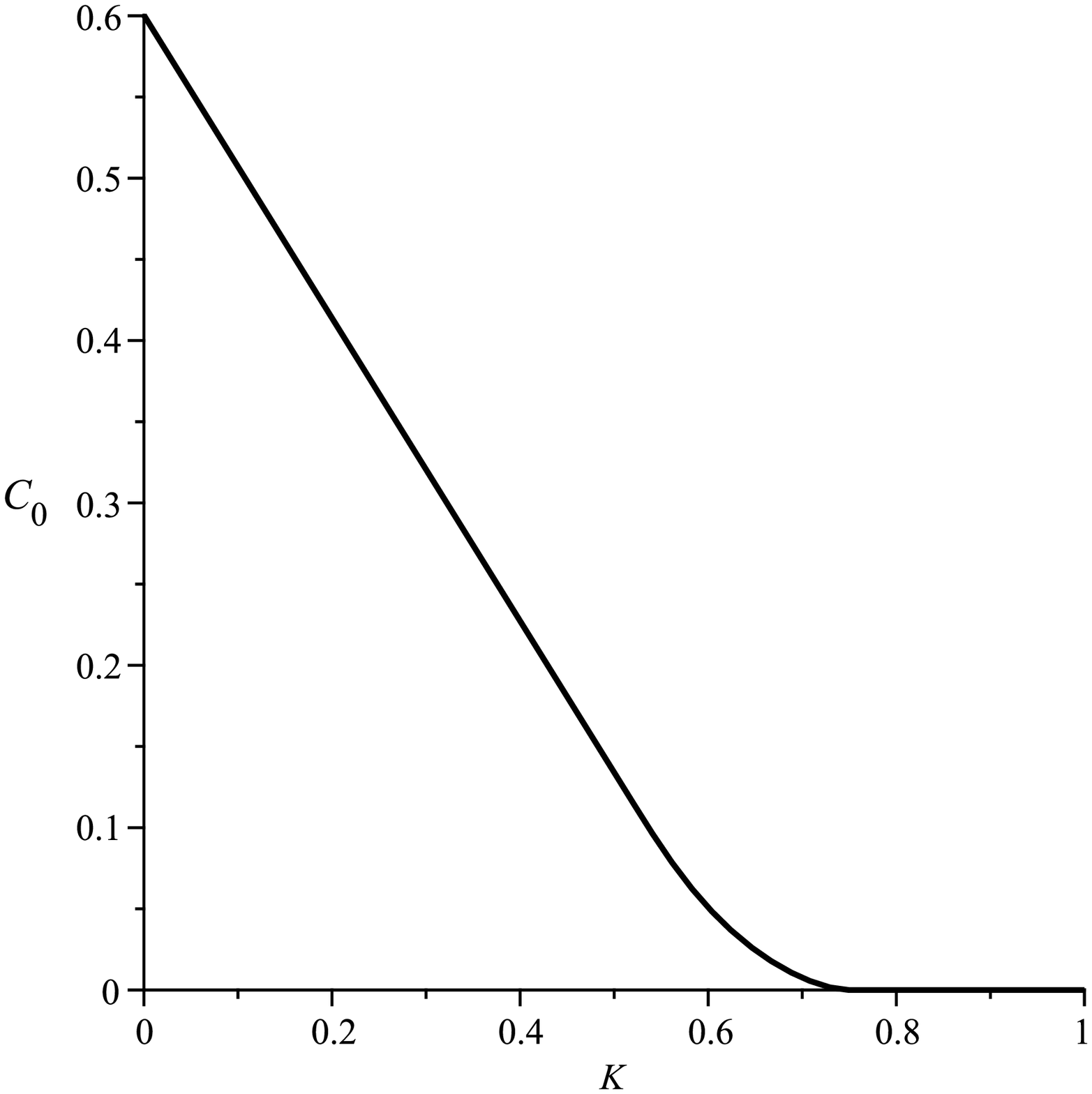}}
\subfloat{\includegraphics[trim=0.5cm 0cm 0cm 0cm, scale=0.31]{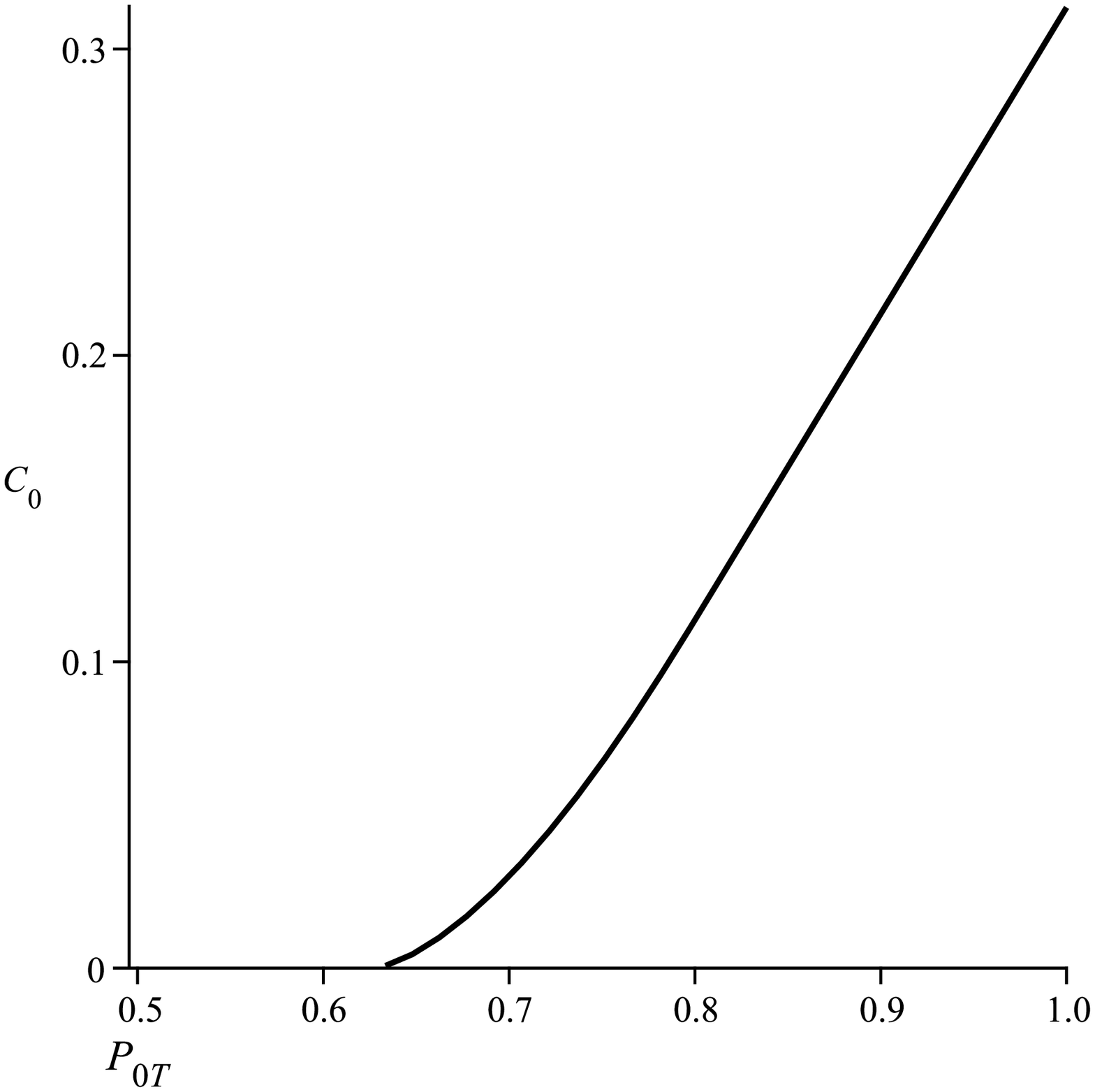}}
\caption{Left panel: the call price as a function of the strike for fixed bond maturity ($T=10$) and 
fixed option maturity ($t=3$), in the second-order coherent chaos model with $\phi^2(s)=(0.1)^{-1} 
\exp(-0.1 s)$. Right panel: the call price as a function of the initial bond price, where the bond price 
is varied by changing its mathrity $T$ from 100 to 5. Other parameters are set as $t=5$, $K=0.7$, 
and $\phi^2(s)=(0.03)^{-1} \exp(-0.03 s)$.}
\label{fig:1} 
\end{figure}

We see, therefore, that in a coherent chaos model of the second order, the 
pricing of options on discount bonds, as well as the determination of the 
associated hedge portfolio, 
are entirely tractable. As an illustrative example, let us consider the 
case in which we choose $\phi^2(s) = \lambda^{-1} \exp(-\lambda s)$ for some 
constant $\lambda$. As indicated above, in a coherent 
chaos model, this is all we need in order to calculate both initial bond prices and call prices. In Figure~\ref{fig:1} on 
the left panel we show the relation between the call price and its strike. Here, the bond maturity is fixed at 
$T=10$ and the option maturity at $t=3$, and we set $\lambda=0.1$. On the right-side of Figure~\ref{fig:1} 
we show how the call price changes with respect to the bond values for different maturities. Here, the option 
maturity has been chosen to be $t=5$, the bond maturities vary from 100 to zero, the strike has been fixed 
at $0.7$, and we have set $\lambda=0.03$.

\subsection{Pricing swaptions in second-order coherent chaos models}

It turns out that swaptions can also be valued in closed form, in a manner analogous to the factorisable 
second-order chaos models of Hughston \& Rafailidis (2004). In this case, the payout of the contract 
can be expressed as follows:  
\begin{eqnarray}
H_{t}^{(2)} = \left(1-P_{t{T_n}}^{(2)}-K \sum_{i=1}^n P_{tT_i}^{(2)} \right)^+,
\end{eqnarray}
where $T_i, \; i=1, \ldots, n$, are specified future payment dates. The initial value of the swaption is then 
given by the expectation: 
\begin{eqnarray}
H_{0}^{(2)} &=& 2\, \mathbb{E}\left[\pi_{t}^{(2)} \left(1-P_{t{T_n}}^{(2)}-K \sum_{i=1}^n P_{tT_i}^{(2)} 
\right)^+\right] \nonumber \\ &=&   2 \, \mathbb{E}\left[\left(A Z_t^2 + B \right)^+\right], 
\label{swaption}
\end{eqnarray}
where $Z_t=R_t/\sqrt{Q_t}$ is defined as before. In the present case, however, we have 
\begin{eqnarray}
A = Q_t[(Q_{T_n} - Q_t) - K \sum_{i=1}^n (1-Q_{T_i})] 
\end{eqnarray} 
and 
\begin{eqnarray} 
B = \half ({Q_{T_n}}- Q_t)^2 + K Q_t\sum_{i=1}^n (1-{Q_T}_i)-\frac{K}{2}\sum_{i=1}^n 
(1-Q^2_{T_i}).
\end{eqnarray}
In other words, the valuation of a swaption proceeds exactly in the same manner as that of a call option, 
except that the coefficients defined as $A$ and $B$ are a little more complicated. Writing $z_1$ and 
$z_2$ for the roots of $Az^2+B=0$, we thus have the following: 

\begin{itemize} 

\item[(i)] When $A<0$ and $B\leq0$, the swaption is evidently worthless.
\item[(ii)] If $A>0$ and $B\geq0$, then we have 
\begin{eqnarray}
H_{0}^{(2)} = P_{0t}^{(2)} - P_{0{T_n}}^{(2)} - K \sum_{i=1}^n P_{0T_i}^{(2)} . 
\end{eqnarray}
\item[(iii)] If $A<0$ and $B>0$ so that $-B/A>0$, then we have 
\begin{eqnarray}
H_{0}^{(2)} &=& 2(A+B)\left[N(z_2)- N(z_1)\right]+ 2 A \left[ z_1 \rho(z_1)- z_2 \rho(z_2)\right] \nonumber \\ 
&=&  \left( P_{0t}^{(2)} - P_{0{T_n}}^{(2)} - K \sum_{i=1}^n P_{0T_i}^{(2)}\right)\left[1 - 
2 N(z_1)\right]+ 4 A z_1 \rho(z_1).
\end{eqnarray}
\item[(iv)] If $A>0$ and $B< 0$ so that $-B/A>0$, then we have 
\begin{eqnarray}
H_{0}^{(2)} = 2 N(z_1) \left( P_{0t}^{(2)} - P_{0{T_n}}^{(2)} - K \sum_{i=1}^n P_{0T_i}^{(2)}\right) - 
4 A z_1 \rho(z_1).
\end{eqnarray}
\end{itemize}

\subsection{Bond option pricing for third-order coherent chaos models} 

The pricing kernel for a third-order pure coherent chaos model can be 
expressed in the form 
\begin{eqnarray}\label{pi3}
\pi_{t}^{(3)} = 6 (1-Q_t)X_{t}^{(4)}+ (1-Q_t^2)X_{t}^{(2)} + \frac{1}{6}(1-Q_t^3),
\end{eqnarray}
and consequently the bond price process is given by 
\begin{eqnarray}
P_{tT}^{(3)}=\frac{36(1-Q_T)X_{t}^{(4)}+6(1-Q_T^2)X_{t}^{(2)} + (1-Q_T^3)}
{36(1-Q_t)X_{t}^{(4)}+6(1-Q_t^2)X_{t}^{(2)} + (1-Q_t^3)}.
\label{eq:65} 
\end{eqnarray}
The value at time zero of a European-style call option on a discount bond in 
this model is thus expressible as 
\begin{eqnarray} 
C_{0}^{(3)} &=& \mathbb{E}\left[\frac{\pi_{t}^{(3)}}{\pi_{0}^{(3)}}\left( 
P_{tT}^{(3)}-K\right)^+\right] \nonumber \\ &=& 3!\; \mathbb{E}\left[ 
\left(AZ_t^4 + BZ_t^2 +C\right)^+\right], 
\end{eqnarray}
where we have defined 
\begin{eqnarray}
A = \frac{1}{4} Q_t^2 \left[(1- Q_T)-K(1-Q_t)\right], \label{A3}  
\end{eqnarray} 
\begin{eqnarray} 
B =  \frac{1}{2} Q_t \left[(1- Q_T^2) - K(1- Q_t^2)\right] - \frac{3}{2} Q_t^2 \left[(1-Q_T) - 
K(1-Q_t)\right], \label{B3} 
\end{eqnarray}
and 
\begin{eqnarray} 
C &=& \frac{1}{6}\left[(1-Q_T^3)-K(1-Q_t^3)\right] - \frac{1}{2} Q_t \left[(1-Q_T^2)-
K(1-Q_t^2)\right] \nonumber \\ && \quad +\frac{3}{4}
Q_t^2 \left[ (1-Q_T)-K(1-Q_t)\right].
\end{eqnarray}
The roots of the quartic polynomial $A z^4 +B z^2 +C$ can easily be obtained: 
\begin{eqnarray}
z_1 = -\sqrt{\frac{-B-\sqrt{\delta}}{2A}}, \quad z_{2}=-
\sqrt{\frac{-B+\sqrt{\delta}}{2A}},  \quad 
z_{3}= -z_2, \quad  z_{4}= -z_1,
\end{eqnarray}
where $\delta=B^2-4AC$. Thus, we have the following: 

\begin{itemize}
\item[(i)] If $A=0$ then $B= \frac{1}{2} Q_t (1-Q_T)(Q_T - Q_t)$ and 
$C= \frac{1}{6}(1-Q_T)(Q_T-Q_t)(Q_T -Q_t + 1 - Q_t)$. Since $C \geq 0$ the option is 
always in the money and we have
\begin{eqnarray}
C_{0}^{(3)} &=& \frac{6}{\sqrt{2 \pi}} \int_{-\infty}^\infty (Bz^2+C)\re^{-\frac{1}{2}z^2} \rd z %\\ &=& 6(B+C) 
\nonumber \\ &=& (1-Q_T) (Q_T -Q_t) ( 1+Q_t +Q_T).
\end{eqnarray}

\item[(ii)] If $A>0$ and $-B-\sqrt{\delta}\geq 0$ then this implies that 
$B \leq -\sqrt{\delta}$. For $C \leq 0$, we have that . It is necessary that $C>0$, since there are four roots and the call option in this case is
\begin{eqnarray}
C_0^{(3)} &=& \frac{3!}{\sqrt{2 \pi}} \left(\int_{-\infty}^{z_2} + \int^{z_1}_{z_4} +\int^{\infty}_{z_3}  \right)
(Az^4 + Bz^2 + C) \; {\re}^{-\frac{1}{2}z^2} \rd z \nonumber \\ &=& 6(3A + B +C) (2N(z_1)+2N(z_2)-1)\nonumber \\ && - 12(3A+B)  (z_2 \rho(z_2) + z_1 \rho(z_1)) \nonumber \\ && - 12A( z^3_2 \rho(z_2) + z_1^3 \rho(z_1)) .
\end{eqnarray}

\item[(iii)] If $A>0$ and $-B-\sqrt{\delta}<0$ but $-B+\sqrt{\delta} \geq 0$ then the 
initial value of the call option is 
\begin{eqnarray}
C_0^{(3)} = \frac{3!}{\sqrt{2 \pi}} \left(\int_{-\infty}^{z_2} + \int^{\infty}_{z_3}  \right)
(Az^4 + Bz^2 + C) \; {\re}^{-\frac{1}{2}z^2} \rd z .
\end{eqnarray}
Performing the integration and noting that $z_2 = - z_3$, we find that
\begin{eqnarray}
C_0^{(3)} = 12(3A + B + C) N(z_2) - 12 (Az_2^2 + 3A+ B)z_2 \rho(z_2). 
\end{eqnarray}

\item[(iv)] If $A>0$ and $-B+\sqrt{\delta}<0$ then $C<0$ and the value of the option 
is zero.

\item[(v)] If $A<0$ and $-B- \sqrt{\delta}> 0$ then
\begin{eqnarray}
C_0^{(3)} &=& \frac{3!}{\sqrt{2 \pi}}\int_{-\infty}^\infty (Az^4 + Bz^2 + C) \; 
{\re}^{-\frac{1}{2}z^2} \rd z \\ &=& 6(3A + B + C) . 
\end{eqnarray}

\item[(vi)] If $A<0$ and $-B+ \sqrt{\delta}> 0$ but $-B- \sqrt{\delta} \leq 0$ then
\begin{eqnarray}
C_0^{(3)} &=& \frac{3!}{\sqrt{2 \pi}}\int_{z_1}^{z_4} (Az^4 + Bz^2 + C) \; 
{\re}^{-\frac{1}{2}z^2} \rd z \\ &=& 6(3A + B + C)(1-2N(z_1)) + 12 z_1 \rho(z_1) 
(A z_1^2 + 3A +B). 
\end{eqnarray}

\item[(vii)] Finally, if $A<0$ and $-B+ \sqrt{\delta}\leq 0$ then
\begin{eqnarray}
C_0^{(3)} &=& \frac{3!}{\sqrt{2 \pi}}\left( \int_{z_1}^{z_2} + \int_{z_3}^{z_4}  \right) 
(Az^4 + Bz^2 + C) \; {\re}^{-\frac{1}{2}z^2} \rd z \nonumber \\ 
&=& 12(3A + B + C)(N(z_2)-N(z_1)) + 12 z_1 \rho(z_1) (A z_1^2 + 3A +B) 
\nonumber \\ && - 12 z_2 \rho(z_2) (A z_2^2 + 3A +B).
\end{eqnarray}

\end{itemize}

In summary, we observe that in general for a coherent chaos model of order $n$ 
the pricing kernel is just a polynomial of order $2n-2$ in the single Gaussian process 
$\{R_t\}$. It follows that for the valuation of an option or a swaption, the relevant 
calculation reduces to that of taking the expectation of the positive part of a polynomial 
of the same order in the standard normal random variable $Z_t$. In other words, the 
problem reduces to the identification of the roots of a polynomial of order $2n-2$. 
Since such an elementary root-finder can easily be carried out numerically, we 
therefore find that semi-analytic expressions for both option and swaption prices are 
always available in the case of an $n^{\rm th}$-order coherent chaos model.

\section{Incoherent chaos models}
\label{sec:7}

We have examined the coherent chaos interest-rate models in some detail with 
the view to generalise them to more generic cases, which we might call \textit{ 
incoherent chaos models}. As indicated above, our key observation is that an 
arbitrary ${\mathcal F}_\infty$-measurable square-integrable random variable 
$X_{\infty}^{(n)}$ associated with an $n^{\rm th}$-order chaos expansion can be 
expressed as a linear combination of the `coherent' log-normal random variables 
$X_\infty^{(n)}(\phi_i)$ for different structure functions $\phi_i(s)$. Putting the matter 
differently, we have the representation 
\begin{eqnarray}
X_{\infty}^{(n)} = \sum_{i} c_i \,X_{\infty}^{(n)}(\phi_i) , 
\label{eq:81}
\end{eqnarray}
where for clarity we have written  
\begin{eqnarray}
X_{\infty}^{(n)}(\phi_i) = 
\int_0^\infty \!\! \int_0^{s_1} \!\!\!\cdots\! \int_0^{s_{n-1}} \phi_i(s_1) \phi_i(s_2) 
\cdots \phi_i(s_n) \, \rd W_{s_n} \!\cdots \rd W_{s_2} \rd W_{s_1} 
\label{eq:82}
\end{eqnarray}
for each $i$ for the coherent random variables. The pricing kernel associated with 
an arbitrary $n^{\rm th}$-order chaos model can thus be obtained by calculating: 
\begin{eqnarray} 
\pi_{t}^{(n)} = \mathbb{E}_t\left[\left(\sum_{i} c_i \, 
X_{\infty}^{(n)}(\phi_i)\right)^2\right] - \left(\mathbb{E}_t\left[\sum_{i}
c_i \, X_{\infty}^{(n)}(\phi_i)\right] \right)^2.
\label{eq:83}
\end{eqnarray} 

The second term in the right side of (\ref{eq:83}) can easily be evaluated on account 
of the martingale relation $\mathbb{E}_t[X_{\infty}^{(n)}(\phi_i)]=X_{t}^{(n)}(\phi_i)$. 
Calculating the first term on the right side of (\ref{eq:83}) is evidently a little more 
complicated. We proceed as follows. First, observe from (\ref{eq:82}) that the random 
variable $X_\infty^{(n)}(\phi_i)$ can be written in the following recursive form: 
\begin{eqnarray} 
X_\infty^{(n)}(\phi_i) = \int_0^\infty \phi_i(t) X_t^{(n-1)}(\phi_i) \rd W_t . 
\label{eq:84}
\end{eqnarray} 
It follows, by use of the conditional form of the Wiener-Ito isometry, that 
\begin{eqnarray} 
\mathbb{E}_t \left[ X_{\infty}^{(n)}(\phi_i) X_{\infty}^{(n)}(\phi_j) \right] &=& 
X_{t}^{(n)}(\phi_i) X_{t}^{(n)}(\phi_j) \nonumber \\ && \hspace{-1.0cm} 
+ \int_t^\infty \phi_i(u) \phi_j(u) {\mathbb E}_t 
\left[ X_{u}^{(n-1)}(\phi_i) X_{u}^{(n-1)}(\phi_j) \right] \rd u . 
\label{eq:85}
\end{eqnarray} 
An application of the recursion relation (\ref{eq:84}) on $X_{u}^{(n-1)}$ together 
with another use of the conditional form of the Wiener-Ito isometry then shows that what remains in the conditional expectation on the right side of (\ref{eq:85}) is reduced to order $n-2$. By iteration, 
we then find that 
\begin{eqnarray}
\mathbb{E}_t \left[ X_{\infty}^{(n)}(\phi_i) X_{\infty}^{(n)}(\phi_j) \right] = \sum_{k=0}^{n} 
\langle \phi_i,\phi_j\rangle_t^{(k)} X_{t}^{(n-k)}(\phi_i) X_{t}^{(n-k)}(\phi_j) , 
\label{eq:86}
\end{eqnarray} 
where for simplicity of notation we have written 
\begin{eqnarray}
\langle \phi_i,\phi_j\rangle_t^{(k)}  
= \int\limits_t^\infty \!\! \int\limits_t^{s_1} \cdots\! \int\limits_t^{s_{k-1}} 
\phi_i(s_1) \phi_j(s_1) \cdots \phi_i(s_k) \phi_j(s_k) \, \rd s_k \cdots \rd s_1 ,
\label{eq:87}
\end{eqnarray} 
with $\langle \phi_i,\phi_j\rangle_t^{(0)} = 1$. Putting these together, we thus deduce the 
following expression: 
\begin{eqnarray}
\pi_t^{(n)} &=& \sum_{i,j} c_i c_j \! \left[ \sum_{k=0}^{n} 
\langle \phi_i,\phi_j\rangle_t^{(k)} X_{t}^{(n-k)}(\phi_i) X_{t}^{(n-k)}(\phi_j) - 
X_{t}^{(n)}(\phi_i) X_{t}^{(n)}(\phi_j) \right]  \nonumber \\ &=& 
\sum_{i,j} c_i c_j \sum_{k=1}^{n} 
\langle \phi_i,\phi_j\rangle_t^{(k)} X_{t}^{(n-k)}(\phi_i) X_{t}^{(n-k)}(\phi_j)
\end{eqnarray} 
for the pricing kernel in the incoherent model. \\ 

\noindent \textit{Example 1}: As an illustrative example, consider the case in which $n=2$, 
and suppose that the chaos expansion involves the superposition of a pair of functions: 
\begin{eqnarray} \label{phi ex inc}
\Phi(s_1,s_2)= \sum_{i=1}^2 c_i  \; \phi_i(s_1) \; \phi_i(s_2) . 
\end{eqnarray}
In this case, after some elementary calculation, we obtain the following expression for 
the pricing kernel:
\begin{eqnarray}\label{incoherent pi}
\pi_{t}^{(2)}  &=& 2 c_1^2 \left( R_t^2 \left(\phi_1)(1-Q_t(\phi_1)\right) + \half\left(1-Q_t(\phi_1)\right)^2 
\right) \nonumber \\ &&  + 2c_2^2 \left( R_t^2 \left(\phi_2)(1-Q_t(\phi_2)\right)+ \half\left(1-Q_t(\phi_2) 
\right)^2   \right) \nonumber \\ && + 2 c_1 c_2 R_t(\phi_1) R_t(\phi_2) 
\int_t^\infty \phi_1(s) \phi_2 (s) \rd s \nonumber \\ && + c_1 c_2 \left( \int_t^\infty \phi_1(s) \phi_2 (s) 
\rd s \right)^2  .
\end{eqnarray} 
Here, we have written $R_t(\phi_i)=\int_0^t \phi_i(s)\rd W_s, \; i=1,2,$ for the two Gaussian processes. 
We see therefore that the pricing kernel in this case is a simple quadratic cross-polynomial of 
the two Gaussian state variables $\{R_t(\phi_1)\}$ and $\{R_t(\phi_2)\}$. Analogous results 
naturally hold for higher $n$ and for a larger number of terms in the expansion. \\

\noindent \textit{Example 2}: More generally, we may consider an `incoherent' model that consists of 
a combination of coherent terms of different chaos orders. As an illustrative example, suppose that 
$X_\infty$ is given by a combination of the first order and an $n^{\rm th}$ order coherent chaos element: 
\begin{eqnarray}
X_\infty = X_\infty^{(1)}(\phi_1) + X_\infty^{(n)}(\phi_2) . 
\end{eqnarray}
Taking the conditional variance and making use of (\ref{eq:86}), the associated pricing kernel takes the 
following form: 
\begin{eqnarray}
\pi_t^{(n)} = 1 - Q_t(\phi_1) + \! \sum_{k=1}^n \frac{\left[X_t^{(n-k)}(\phi_2)\right]^2}{(k!)^2} 
+ 2 X_t^{(n-1)}(\phi_2) \! \int\limits_t^\infty \!\! \phi_1(s) \phi_2(s) \rd s,
\end{eqnarray}
from which the corresponding term-structure dynamics can easily be derived.

\section{Finite dimensional realisations of coherent chaos models}
\label{sec:8}

In this section we consider the case in which the Hilbert space of square-integrable 
functions is approximated by (or replaced with) a finite-dimensional Hilbert space. To 
facilitate the calculation, we make use of the Dirac delta function so that the function 
$\phi(x)$ is given by the square root of a weighted sum of a finite number of delta 
functions. The idea of finite-dimensional realisations is to generate a set of models 
that can be calibrated purely in terms of a finite number of available market data, 
without evoking the assumption of hypothetical initial bond prices across all maturities 
that do not exist. 

To proceed, we recall first that we have made the normalisation convention such that 
\begin{eqnarray}
Q_\infty = \int_0^\infty \phi^2(s) \rd s = 1. 
\end{eqnarray}
Hence, $\phi^2(s)$ can be interpreted as defining a probability 
density function, except that now in effect in a finite-dimensional space. More precisely, and 
strictly speaking, we continue to work in the infinite-dimensional Hilbert space, but by 
facilitating a finite number of distributions rather than functions, the analysis in effect reduces 
to that based on a finite-dimensional Hilbert space. We therefore choose $\phi^2$ to take the 
following form: 
\begin{eqnarray}
\phi^2(s)=\sum_{i=1}^{N+1} p_{i}  \delta (s-T_i), 
\label{eq:94} 
\end{eqnarray}
where $\delta(s)$ denotes the standard Dirac delta function, and the $\{T_i\}_{i=1,\ldots,N}$ 
represent different maturity dates of bonds for which prices are available in the market; $N$ 
indicates how many of these there are. The coefficients $\{p_{i}\}$ are probability weights so 
that $0 \leq p_{i} \leq 1$ and that $\sum_{i=1}^{N+1} p_{i} = 1$. For this choice of $\phi$ the 
integral $Q_{t}$ then takes the following form of a piecewise step function: 
\begin{eqnarray}\label{Q fdr}
Q_{t} = \sum_{i=1}^{N+1} p_{i} \mathds{1}_{\{T_{i} \leq t \}} 
= \sum _{i=1}^{N+1} \mathds{1}_{\{T_{i} \leq t < T_{i+1} \}} \;\sum_{j=1}^i p_{j}. 
\end{eqnarray}
We remark that in (\ref{eq:94}) and (\ref{Q fdr}) 
we have introduced an arbitrary $T_{N+1}>T_N$ such that the bond price is assumed 
to become zero at this point. Note that $T_{N+1}$ is not an expiry date, but rather an 
arbitrary time that is beyond the maturity date of the bond in the market with the longest 
lifetime. The choice of $T_{N+1}$ will not affect the valuation of contracts whose 
maturities are $\leq T_N$; thus the analysis below will not be affected by the arbitrariness 
of the choice of $T_{N+1}$. The reason for introducing $T_{N+1}$ is merely to fulfil the 
asymptotic condition (\ref{lim P}) of axiom (d) in the finite-dimensional setup. Conversely, 
had we not introduced $T_{N+1}$, the discussion below would not be affected. 

Let us now consider the $n^{\rm th}$-order coherent chaos model associated with the 
positive square-root of (\ref{eq:94}). A short calculation using expression (\ref{Q fdr}) in the 
result (\ref{P0t}) for $P_{0t}^{(n)}$ shows that the corresponding 
initial bond prices with maturity $t$ admit the following representation: 
\begin{eqnarray} 
P_{0t}^{(n)} = 1 -  \sum _{i=1}^{N+1} \mathds{1}_{\{T_{i} \leq t < T_{i+1} \}} \; 
\left(\sum_{j=1}^i p_{j}\right)^n. 
\label{eq:96}
\end{eqnarray}

\begin{figure}[h!]
\centering
%\begin{center}
\includegraphics[scale=0.42]{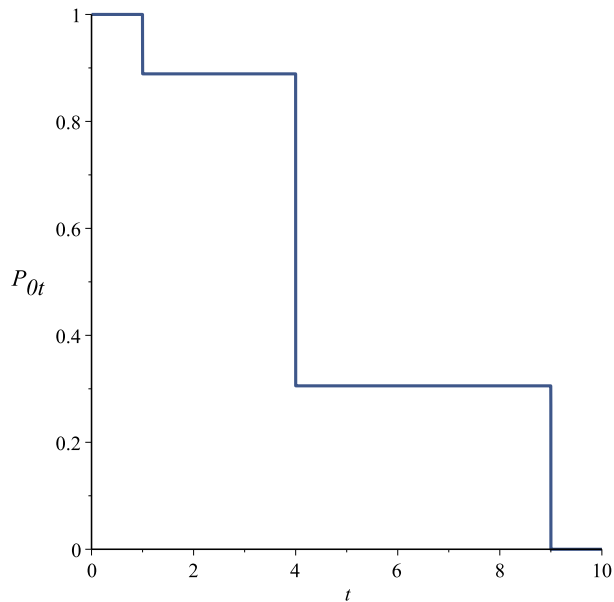}
\caption{Initial bond prices $P_{0t}^{(n)}$ as a function of maturity time for $n=2$. Two 
bond prices are assumed given at times $T_1 = 1$ and $T_2 = 4$; while at time $T_3=9$ 
the bond price is assumed to jump to zero. The parameter values chosen here are 
$p_1 = \frac{1}{6}$, $p_2 = \frac{1}{2}$, and $p_3 = \frac{1}{3}$.}
%\end{center}
\label{fig:2} 
\end{figure}

To illustrate the initial term structure in the present model we sketch in Figure~\ref{fig:2} 
an example of the initial bond prices as a function of the maturity time for $n=2$. Three 
maturities have been used here: $T_1=1$, $T_2=4$ and the `artificially' chosen $T_3=9$. 
According to (\ref{eq:96}) the initial value $P_{0T_1}$ of a bond expiring at time $T_1$ 
(where the first jump occurs) is $P_{0T_1}=1-p_{1}^n$, the price at time $T_2$ is 
$P_{0T_2} = 1-(p_{1}+p_{2})^n$, and so on. Hence in the pure second-order coherent chaos 
model with only two data points specified, we have the values $P_{0T_1} = 1-p_{1}^2$, 
$P_{0T_2} = 1-(p_{1}+p_{2})^2$, and $P_{0T_3} = 1-(p_{1}+p_{2}+p_{3})^2=0$. 

\begin{figure}[t!]
\centering
\includegraphics[scale=0.3]{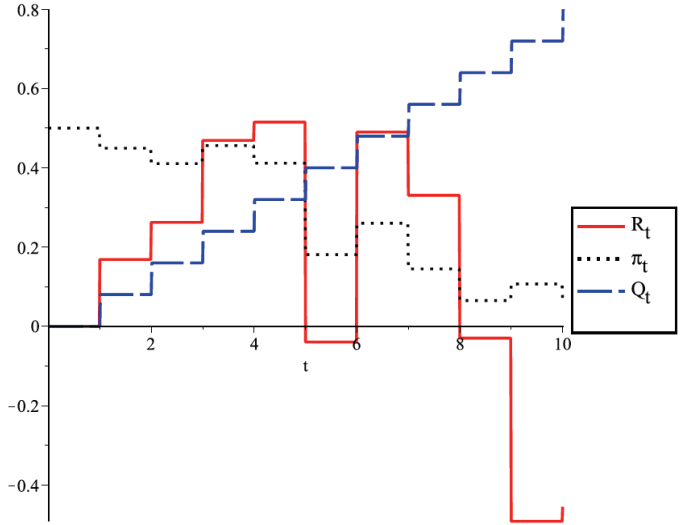}
\caption{Sample path simulations for the Gaussian driver $\{R_t\}$, its quadratic variation 
$\{Q_t\}$, and the resulting pricing kernel $\{\pi_{t}\}$, in the case of a finite-dimensional 
coherent chaos model with $n=2$. The parameters chosen here are $p_i = 0.08$ 
and $T_i=i$ for $i=1 \ldots 10$, and $p_{11}=0.2$ with $T_{11}>10$ arbitrary.}
\label{fig:3} 
\end{figure}

\begin{figure}[t!]
\centering 
\includegraphics[scale=0.3]{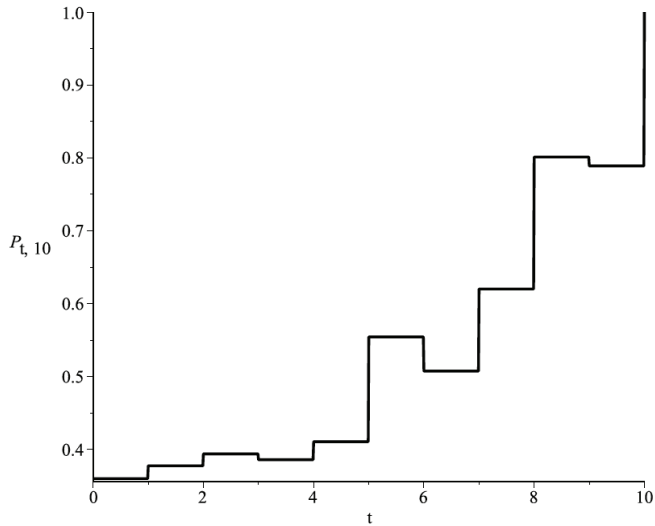}
\caption{A sample path simulation for the discount bond price process $\{P_{tT}\}$ in the 
coherent chaos model for order $n=2$. The parameters chosen here as $p_i=0.08$ 
for $i=1,\ldots,10$, and the bond maturity is set as $T=10$. The bond price fluctuates up 
and down, but eventually converges to its terminal value one. The prices jump at
the points at which the initial bond prices are assumed given in accordance with (\ref{eq:96}). }
\label{fig:4} 
\end{figure}

Since it is assumed that the maturities $\{T_i\}_{i=1,\ldots,N}$ are the points where we have market data 
for the bond prices, we see that the probability weights $\{p_i\}$ can be calibrated from the 
initial market prices of the discount bonds. This in turn determines the function $\phi(s)$, 
which in turn determines the subsequent dynamics for the term structure. Let us proceed 
to analyse the term structure dynamics in the finite-dimensional models. To begin, recall 
that $R_{t} \sim \mathcal{N}(0,Q_{t})$, so in order to determine the Gaussian process 
$\{R_t\}$ whose probabilistic characteristics are identical to that of $\int_0^t \phi(s) \rd W_s$, 
where $\phi(s)$ is of the form (\ref{eq:94}), let us consider a family of independent Gaussian 
random variables  
\begin{eqnarray}
n_i \sim \mathcal{N}\left(0,\sum_{j=1}^i p_{j}\right) \qquad \mbox{for}\; i=1,\ldots,N+1.
\end{eqnarray}
Then we can represent the Gaussian process $\{R_t\}$ according to
\begin{eqnarray} 
R_t = \sum _{i=1}^{N+1} \mathds{1}_{\{T_{i} \leq t < T_{i+1} \}} \;n_i 
\label{eq:98}
\end{eqnarray}
where equality here holds of course in probability. By making use of (\ref{eq:39}) and 
(\ref{eq:41}), the corresponding pricing kernel can be identified in the finite-dimensional 
models. In figure~\ref{fig:3} we illustrate typical sample paths of $\{Q_t\}$, $\{R_t\}$, and 
$\{\pi_{t}^{(2)}\}$ for $n=2$ and $N=10$. For simplicity we have chosen a uniform 
probability and equal spacing: $p_i=0.08$ and $T_i=i$ for $i=1, \ldots,10$. Similarly, in 
figure~\ref{fig:4}, we show a sample path of a corresponding bond price process, where 
we have again chosen $p_i=0.08$ for $i=1, \ldots, 10$ with bond maturity $T=10$.

\begin{figure}[t!]
\centering 
\includegraphics[scale=0.4]{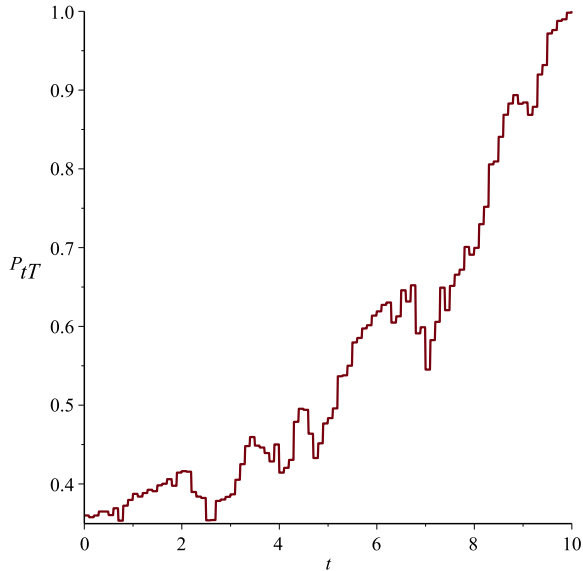}
\caption{A sample path simulation for the discount bond price process $\{P_{tT}\}$ in the 
coherent chaos model for order $n=2$. The parameters chosen here are $p_i=0.008$ 
for $i=1,\ldots,100$, and the bond maturity is set as $T=10$. The bond price fluctuates up 
and down, but eventually converges to its terminal value one. The prices jump at
the points at which the initial bond prices are assumed given in accordance with (\ref{eq:96}). }
\label{fig:5} 
\end{figure}

Summing up the section, we see that using the finite-dimensional approach, a class of 
elementary, highly tractable and easily calibrated models can be constructed. These models have the 
characteristic feature that the discount bond process is piecewise flat and that the 
distribution of the bond price at any time is determined by a ratio of polynomials of 
standard normal random variables. The bond price processes emerging in these models 
can alternatively be viewed as representing `simple' approximations to more sophisticated 
continuous processes. To illustrate how a typical bond price in these models behaves, in 
figure~\ref{fig:5} we show an example of the bond price dynamics where the `grid size' is 
made ten times finer than the one sketched in figure~\ref{fig:4}. It is interesting to note in 
this connection that although the Gaussian process $R_t=\int_0^t \phi(s) \rd W_s$ 
formally appears to be continuous, owing to the appearance of distributions in (\ref{eq:94}) 
the resulting process contains jumps, as shown in figure~\ref{fig:3}.

\section{Conclusion and discussion} 
\label{sec:9}

The purpose of the present paper is to offer an in-depth analysis of the coherent 
chaos interest-rate models of each order with the view that they form the basis 
for generic (incoherent) interest-rate models. We have found that for a pure 
$n^{\rm th}$ order coherent chaos model, the pricing kernel is given by a polynomial 
of order $2n-2$ of a Gaussian state variable. This leads to fairly simple expressions 
for the bond price, the risk premium, and the short rate processes. Additionally, we 
have shown that in all these models it is possible to derive either analytic or 
semianalytic formulae for the pricing of both bond options and swaptions, involving 
at most numerical determination of roots of basic polynomials. 

Single-factor coherent chaos models, although are themselves considerably richer 
than elementary Gaussian interest rate models, are nevertheless too restrictive on 
account of the fact that each coherent chaos model depends on just a single 
functional degree of freedom. For more realistic models, however, it suffices to 
take linear superpositions of coherent chaos elements in the sense described in 
Section~\ref{sec:4}. The resulting expression for the pricing kernel, in the most 
generic case, is somewhat cumbersome, although is nevertheless tractable. 
However, for practical purposes it seems sufficient to consider just a small number 
of (two or three) coherent vectors so as to generate rich and flexible interest-rate 
models. We hope that the results presented here will form a foundation for further 
investigation into this direction. 

By means of comparison, we draw attention to the fact that implementation of chaos 
models has recently been carried out in Grasselli \& Tsujimoto (2011) in the 
case of a third-order chaos model, with the choices $\phi(s_1) = \alpha(s_1)$, 
$\phi(s_1,s_2)= \beta(s_1)$, and $\phi(s_1,s_2, s_3) = \gamma(s_1)$. Their model 
has been shown to fit the forward curve more closely, and with less parameters, as 
compared to currently preferred models employed in the industry. An interesting 
extension would therefore be to consider the implementation of the following 
generalisation: $\phi(s_1) = \alpha(s_1)$, $\phi(s_1,s_2)= \beta(s_1)\beta(s_2)$, 
and $\phi(s_1,s_2, s_3) = \gamma(s_1)\gamma(s_2)\gamma(s_3)$, in line with the 
foregoing discussion, and examine how well the generalised model fits both the 
forward and volatility curves.

In relation to the analysis on finite-dimensional models, it is worth making the following 
observation. For simplicity, if we set $N=1$, then formally we are led to an 
expression of the form 
\begin{eqnarray}
R_t = \int_0^t \sqrt{\delta(s-T_1)} \, \rd W_s 
\label{eq:102}
\end{eqnarray}
for the Gaussian process $\{R_t\}$, the meaning of which \textit{a priori} is not easily 
interpreted. In the present paper we have circumvented the direct handling of processes 
of the form (\ref{eq:102}) and its generalisations by means of identifying for each $t$ an 
alternative random variable whose probability law is identical to that of $R_t$, and used 
this alternative representation to characterise interest-rate dynamics. For a more direct 
analysis of stochastic integrals of the form (\ref{eq:102}), calculus on the multiplication of 
distributions due to Colombeau (1990) might prove useful. We defer such an analysis to 
another occasion.



\begin{thebibliography}{}
\bibitem{Bj1}Bj\"ork, T.: \textit{Arbitrage Theory in Continuous Time}, 3rd ed. 
Oxford: Oxford University Press (2009) 

\bibitem{Label13} Bj\"ork, T.: Topics in interest rate theory. In \textit{Handbooks in Operations Research and Management Science}, J.~R.~Birge \& V.~Linetsky, eds. \textbf{15}, 377-435 (2007)

%\bibitem{Label20} Brody D.C., Hughston L.P., Interest rates and information geometry, Proceedings of the 
%Royal Society, \textbf{457}, 1343-1363 (2001).  

%\bibitem{Label2} Brody D.C., Hughston L.P., Entropy and Information in the Interest Rate Term Structure,
%Quantitative Finance, \textbf{2}, 70-80 (2002).

\bibitem{BM} 
Brigo, D., Mercurio, F.: Interest Rate Models-Theory and Practice: with Smile, Inflation and Credit. Springer (2007)


\bibitem{Label1} Brody, D.C., Hughston, L.P.: Chaos and coherence: a new framework for interest rate modelling. Proceedings of the Royal Society. \textbf{460}, 85-110 (2004)

\bibitem{Label25} Brody, D.C., Hughston, L.P., Mackie, E.: General theory of geometric L\'{e}vy models for dynamic asset pricing. Proceedings of the Royal Society. \textbf{468}, 1778- 1798 (2012)

\bibitem{CAI} Cairns, A.J.: Interest Rate Models: an Introduction (Vol. 10). Princeton University Press (2004)

\bibitem{Label26} Cochrane, J.H.: Asset Pricing. Princeton University Press (2005)

\bibitem{22} Colombeau, J.F.: Multiplication of distributions. American Mathematical Society. \textbf{23(2)}, 251-268 (1990)

\bibitem{CT} Carmona, R., Tehranchi, M.R.: Interest Rate Models: an Infinite Dimensional Stochastic Analysis Perspective. Springer (2007)

\bibitem{label3} Di Nunno, G., Oksendal, B., Proske, F.: Malliavin Calculus for Levy Processes with Applications
to Finance. Springer (2009)

\bibitem{LabelF} Filipovic, D.:Term-Structure Models: A Graduate Course. Springer (2009)

\bibitem{LabelFH1} Flesaker, B., Hughston, L. P.: International models for interest rates and foreign exchange. Net exposure. 55-79 (1997)

\bibitem{Label4} Flesaker, B.,Hughston, L.P.: Positive interest. Risk Magazine \textbf{January}. 46-49 (1996)

\bibitem{Label16} Flesaker, B., Hughston, L.P.: Positive interest: an afterword. Risk Magazine. 120-124 (1998)

\bibitem{Label17} Grasselli, M.R., Hurd, T.R.: Wiener chaos and the Cox-Ingersoll-Ross model. Proceedings of the 
Royal Society. \textbf{461}(2054), 459-479 (2005)

\bibitem{Label23} Grasselli, M.R., Tsujimoto T.: Calibration of chaotic models for interest rates. arXiv preprint arXiv:1106.2478 (2011)

\bibitem{HM} Hughston, L.P., Mina, F.: On the representation of general interest rate models as square-integrable Wiener functionals. In: Recent Advances in Financial Engineering 2011 (Y. Muromachi, H. Nakaoka \& A. Takahashi, eds.).
World Scientific Publishing Company (2012) 


\bibitem{Label5} Hughston, L.P., Rafailidis, A.: A chaotic approach to interest rate modelling. Finance and Stochastics.  \textbf{9}, 43-65 (2004)

\bibitem{Label30} Ikeda, N., Watanabe, S.: Malliavin calculus of Wiener functionals and its applications. From local times to global geometry, control and physics. \textbf{150}, 132-178 (1987)

\bibitem{label6} It\^{o}, K.: Multiple Wiener integral. Journal of Mathematical Society Japan. \textbf{3}(1), 157-169 (1951)

\bibitem{JW} James, J., Webber, N.: Interest Rate Modelling. Wiley-Blackwell Publishing Ltd (2000)

\bibitem{label7} Janson, S.: Gaussian Hilbert Spaces. Cambridge University Press (1997)

\bibitem{label28} Jin, Y., Glasserman, P.: Equilibrium positive interest rates: a unified view. The Review of Financial Studies. \textbf{14}(1), 187-214 (2001)

\bibitem{Label31} Malliavin, P.: Stochastic Analysis. Springer-Verlag (1997)

\bibitem{Label29} Meyer, P.A.: Probability and Potentials. Blaisdell Pub. Co.  (Waltham, Mass) (1966)

\bibitem{label8} Nualart, D.: The Malliavin Calculus and Related Topics. Springer-Verlag (1995)

\bibitem{Label10} Rogers, L.C.G.: The potential approach to the term structure of interest rates and foreign exchange rates. Mathematical Finance. \textbf{7}, 157-176 (1997)

\bibitem{Label9} Rogers,~L.C.G.: One for all: The potential approach to pricing and hedging. In \textit{Mathematics in Industry}, \textbf{8}, \textit{Progress in Industrial Mathematics at ECMI 2004},  A.~Di~Bucchianico,  R.~M.~M.~Mattheij,  M.~A.~Peletier (eds), Heidelberg: Springer-Verlag (2006).

\bibitem{Label27} Rutkowski, M.L.: A note on the Flesaker-Hughston model of the term structure of interest rates. Applied Mathematical Finance. \textbf{4}(3), 151-163 (1997)

\bibitem{Label32} Shoutens, W.: Stochastic Processes and Orthogonal Polynomials. Lecture Notes in Statistics. Springer. \textbf{146} (2000)

\bibitem{Label19} 
Wiener, N.: The homogeneous chaos. American Journal of Mathematics. 
\textbf{60}, 897-936 (1938)



\end{thebibliography}
\end{document}